\documentclass[useAMS,usenatbib]{mn2e}
\usepackage{natbib}
\usepackage{graphicx}
\usepackage{savesym}
\usepackage{listings}
\usepackage[intlimits]{amsmath}
\savesymbol{iint}
\usepackage{txfonts}
\usepackage{bm}
\restoresymbol{TXF}{iint}
\usepackage{amssymb}
\usepackage{hyperref}
\usepackage{amsmath}
\usepackage{tabularx} 
\usepackage{color}
\usepackage{soul}

\def\be{\begin{equation}}
\def\ee{\end{equation}}
\def\ba{\begin{eqnarray}}
\def\ea{\end{eqnarray}}
\def\go{\mathrel{\raise.3ex\hbox{$>$}\mkern-14mu
             \lower0.6ex\hbox{$\sim$}}}
\def\lo{\mathrel{\raise.3ex\hbox{$<$}\mkern-14mu
             \lower0.6ex\hbox{$\sim$}}}

\newcommand{\oder}[2]{\frac{d #1}{d #2}}
\newcommand{\pder}[2]{\frac{\partial #1}{\partial #2}}

\newcommand{\Oder}[2]{\frac{D #1} {D #2}}

\title[Relativistic Centrifugal Instability]{Relativistic Centrifugal Instability}

\author[K.N. Gourgouliatos et al.]{{Konstantinos N. Gourgouliatos\thanks{Email: Konstantinos.Gourgouliatos@durham.ac.uk}$^{1,2}$ \& Serguei S. Komissarov$^{2}$}\vspace{0.4cm}\\
\parbox{\textwidth}{$^{1}$Department of Mathematical Sciences, Durham University, Durham,  DH1 3LE, UK,}\\
{$^{2}$Department of Applied Mathematics, University of Leeds, Leeds LS2 9JT , UK, }}

\begin{document}
\maketitle

\begin{abstract}
Near the central engine, many astrophysical jets are expected to rotate about their axis. Further out they are expected to go through the processes of reconfinement and recollimation. In both these cases, the flow streams along a concave surface and hence, it is subject to the centrifugal force.  It is well known that such flows may experience the Centrifugal Instability (CFI), to which there are many laboratory examples. The recent computer simulations of relativistic jets from Active Galactic Nuclei undergoing the process of reconfinement show that in such jets CFI may dominate over the Kelvin-Helmholtz instability associated with velocity shear \citep{GK:2017}. 
In this letter, we generalise the Rayleigh criterion for CFI in rotating fluids to relativistic flows using a heuristic analysis. We also present the results of computer simulations which support our analytic criterion for the case of an interface separating two uniformly-rotating cylindrical flows. We discuss the difference between CFI and the Rayleigh-Taylor instability in flows with curved streamlines.             
\end{abstract}

\date{\today}

\section{Introduction}
 Over the last few decades, various astronomical studies revealed that both relativistic and non-relativistic 
 accreting cosmic objects often produce spectacular collimated outflows. The speeds of these jets range from $100\,km/s$ in the case of jets associated with young stars \citep{BRD:2007}, to almost the speed of light  in the case of jets are associated with  Active Galactic Nuclei (AGN) \citep{Bridle:1984}, micro-quasars \citep{Mirabel:2010} and Gamma Ray Bursts \citep{KZ:2015}. The current models of the astrophysical jet production include a rapidly rotating central object and magnetic fields and predict that these jets are also rapidly-rotating close to their central engines.  Detailed imaging of proto-stellar jets has already detected such rotation \citep{Zapata:2009, Lee:2017}.   Relativistic jets of AGN may develop curved streamlines at much larger distances as well, where they are expected to change their propagation regime from freely expanding to confined by the pressure of external gas \citep{Sanders:1983,PK:2015}.  Such jets may suffer the centrifugal instability ( CFI,  \cite{GK:2017}).

Lord  \cite{Rayleigh:1917} demonstrated that the rotation of an axially symmetric incompressible and inviscid fluid is unstable provided 
\be
      \frac{d\hat{\Psi}}{dR} <0\,,
      \label{eq:IC1}
\ee
where $\hat{\Psi}=(\Omega R^2)^2$, $\Omega$ is the angular velocity, $R$ is the cylindrical radius. \cite{Bayly:88} has shown that the unstable modes can be highly localised near the streamlines where the Rayleigh condition is satisfied, thus increasing the significance of the Rayleigh instability criterion as a local condition. In particular, this condition is always satisfied, at least locally, if $\Omega$ vanishes at some radius. For example, in curved pipe flows the fluid comes to rest within a boundary layer. This leads to the phenomenon of G\"ortler vortices \citep{Gortler:1955}, which may trigger turbulent cascade and disrupt the flow \citep{Saric:1994}. In the case of reconfined jets, a similar configuration emerges because the external medium is at rest and the jet boundary is concave.
  
In the case of astrophysical jets, not only the velocity but also mass density are expected to show significant variation across the jets. In addition, these jets are mostly supersonic and hence one has to allow for fluid compressibility. Finally, the flow speed can be relativistic. In what follows, we use a simple heuristic approach and generalise the Rayleigh criterion to account for these factors. In order to reduce the level of complexity and allow clear-cut conclusions we confine our study to the case of a rotating unmagnetised ideal relativistic fluid.  The analysis is complemented with computer simulations, which focus on the case where the instability is produced by a jump of the physical parameters at a given radius, reflecting the strong contrast between the jet and its environment.   
\\    

\section{ Generalised Rayleigh Criterion} 
The equations of ideal relativistic hydrodynamics include the continuity equation 
\be
      \frac{1}{\sqrt{|g|}} \pder{\sqrt{|g|} \rho u^\nu}{x^\nu} =0 \,,
\label{CE}
\ee
and the energy-momentum equation 
\be
  \frac{1}{\sqrt{|g|}} \pder{\sqrt{|g|} T^{\nu}_{\mu}}{x^\nu} - 
  \frac{1}{2} T^{\alpha\beta} \pder{g_{\alpha\beta}}{x^\mu} =0 \,,
  \label{EME}
\ee
where $T^{\alpha\beta} = w u^\alpha u^\beta + g^{\alpha\beta}$ is the stress-energy-momentum tensor, $g^{\alpha\beta}$ is the metric tensor and $g$  is its determinant, $w=e+p$ is the relativistic enthalpy per unit volume,  $e$ is the internal energy density, $\rho$ is the rest mass density, $p$ is the pressure and $u^\mu$ is the 4-velocity vector of the fluid \citep{LL75}.   These equations are written in terms of coordinate derivatives and involve the components of vectors and tensors as measured in the corresponding non-normalised coordinate basis. 
Combining the two, one obtains the equation of motion of the fluid element  
\be 
\rho \Oder{h u_{\mu}}{\tau} = -\pder{p}{x^\mu}  -\frac{p}{\sqrt{|g|}}\pder{\sqrt{|g|}}{x^\mu}  + 
\frac{1}{2} T^{\alpha\beta} \pder{g_{\alpha\beta}}{x^\mu} \,,  
\label{EMEQ}
\ee
where $h=w/\rho$ is the enthalpy per unit mass and  $D/D\tau =u^\nu \partial/\partial x^\nu$ is the absolute derivative along the world-line of the fluid element.   

Here we consider only axisymmetric flows in Minkowski space-time and employ cylindrical spatial coordinates $\{z,R,\phi\}$ where 
$\partial/\partial\phi$=0). In this case,  the time component of Eq.(\ref{EMEQ}) reads  
\be
    \Oder{(h u_t)}{\tau} = -\pder{p}{t} \,, 
    \label{time-comp}
\ee
the azimuthal component is
\be
    \Oder{(h u_\phi)}{\tau} = 0 \,, 
        \label{phi-comp}
\ee
where $u_\phi = \Gamma\Omega R^2$, and  the radial component reads 
\be
    \rho \Oder{(h u_R)}{\tau} = -\pder{p}{R} + \frac{w u_{\hat{\phi}} ^2} {R} \,.
    \label{r-comp}
\ee 
where $u_{\hat{\phi}} = u_\phi/R$ is the azimuthal component of the 4-velocity in the normalised coordinate basis. 
Eq.(\ref{phi-comp}) states that $hu_\phi$ is an integral of axisymmetric motion (this constitutes the angular momentum conservation) whereas Eq.(\ref{time-comp}) shows that $hu_t$ and hence $L=u_\phi/u_t$ are generally not\footnote{\citet{Seguin75} carried out a similar heuristic derivation of the stability criterion for the problem of rotating relativistic stars, where he used $L$ as an integral of motion. This can only be justified if $\partial p/\partial t =0 $, which is not self-evident. Even if one can make $\partial p/\partial t$ arbitrarily small by employing a  sufficiently slow motion this involves an increase of the travel time and the overall variation of $hu_t$ may remain finite (see Eq.(\ref{time-comp})). In our analysis we do not assume that $L$ is an integral of motion and the Seguin criterion for the case of perfect fluid does not reduce to ours in the limit of Minkowski space-time .}. 
In equilibrium, the radial force vanishes
\be
    f_R =  -\pder{p}{R} + \frac{w u_{\hat{\phi}} ^2} {R} =0 \,.
    \label{eq:balance}
\ee

{\it Discontinuous case:}
 First we consider the stability at the discontinuity between two rotating flows, located at the radius $R=R_d$.  Here we use suffixes  ``1'' and ``2'' to denote the fluid parameters just below and above $R_d$ respectively.   
Following the Rayleigh argument, we consider fluid rings pushed across the discontinuity but instead of computing the corresponding change in their kinetic energy we simply check if they become subject to a restoring force.  If the motion is slow compared to the sound speed then after the crossing the ring adjusts its pressure to that of its new surrounding.  Hence the pushed upwards ring  will experience the force 
\be
    f_R^* =  -\frac{1}{R_d}[w u_{\hat{\phi}}^2] =  -\frac{1}{R_d^3}[\Psi] \,,
\ee   
where  
\be
    \Psi =  w ({u}_{\hat\phi} R)^2   =  w \Gamma^2 (\Omega R^2)^2 
    \label{eq:RRD}
\ee
and $[\Psi]=\Psi_2-\Psi_1$. The force will push the ring further up  provided 
\be
     [\Psi] < 0 \,,
     \label{eq:IC2}
\ee
which is the instability condition for the discontinuity.  The same conclusion holds for the ring pushed downwards.  
In the Newtonian limit $\Psi=\rho(\Omega R^2)^2$. 

 
When $\Omega$ is continuous across the discontinuity\footnote{ The continuity of $\Omega$ does not necessarily imply the continuity of the axial velocity $v^z$ and hence the continuity of $\Gamma$.} the criterion reduces to $[\Gamma^2 w]<0$ which reads $[\rho]<0$ in the Newtonian limit. This special case may be identified with the Rayleigh-Taylor instability \citep[RTI,][]{Rayleigh:1883, Taylor:1950}, where the centrifugal force plays the role of gravity.    

{\it Continuous case:}
Now we turn to the case with continuous variation of parameters. This time a fluid ring is displaced from $R=R_1$ to $R=R_2=R_1+\delta R$. After the displacement the force acting on the ring is 
\be
    df_R^* =  -\frac{1}{R_2} \left( w_2 (u_{2,\hat\phi})^2 - \tilde{w}_1 (\tilde{u}_{1,\hat\phi})^2 \right) \,,
    \label{eq:force}
\ee   
where $\tilde{q}_1 = q_1 +\delta q $ indicates the value of quantity $q$ after the displacement and $q_2$ is the value of this quantity at $R=R_2$ in the equilibrium configuration. For adiabatic motion $\delta e = h \delta\rho$ and $\delta\rho = \delta P/h a^2$, where $a$ is the sound speed. From these equations and Eq.~(\ref{eq:balance}), it follows
that 
\be
     \frac{\delta\rho}{\rho_1} = \frac{u_{1,\hat\phi}^2}{a_1^2} \frac{\delta R}{R_1} 
     \label{var-rho}
\ee
and
\be
    \frac{\delta w}{w_1} =   \left( \frac{1+ a_1^2}{a_1^2} \right) u_{1,\hat\phi}^2 \frac{\delta R}{R_1} \,.
    \label{var-w}
\ee 
Combining these equations with the angular momentum conservation $h_1u_{1,\hat\phi} R_1=\tilde{h}_1\tilde{u}_{1,\hat\phi} R_2$, 
we find 
\be
     \tilde{u}_{1,\hat\phi}= {u}_{1,\hat\phi} \frac{R_1}{R_2} \left( 1- {u}_{1,\hat\phi}^2\frac{\delta R}{R_1} \right) \,.
     \label{var-u}
\ee
Substituting Eqs.(\ref{var-rho}-\ref{var-u})  into Eq.(\ref{eq:force}) and retaining only the first order terms, we find
\be
  df_R^* =  \left(\Psi_1M_1^2- R_1\left.\!\!\oder{\Psi}{R} \right|_{R=R_1}  \right) \frac{\delta R}{R_1^4}\,, 
\ee
 where 
$M=\Gamma \Omega R/ (\Gamma_a a)$ is the relativistic Mach number of the rotational motion. This immediately leads to the local instability condition 
\be
     \oder{\ln\Psi}{\ln R} < M^2 \,.
     \label{eq:IC3}
\ee

When expressed in terms of a finite jump the criterion (\ref{eq:IC3}) is fully consistent with (\ref{eq:IC2}) for a discontinuity.  In the Newtonian limit, the instability criterion has the same form as (\ref{eq:IC3}) but with $M=\Omega R/a$ and 
\be 
      \Psi = \rho (\Omega R^2)^2 \,.
\ee     
Furthermore, in the incompressible limit ($M=0$ and $\rho={\rm const}$) we recover the original Rayleigh criterion.  
 \\
 
\section{Computer Simulations} 
In this section we describe the axisymmetric computer simulations used to verify the instability criterion for the discontinuous case.  To this aim, we consider rotating fluids with initial cylindrical geometry ($\partial/\partial z=0$)  and vanishing axial velocity ($v_z=0$).  Both the density and the angular velocity are piecewise constant: 
\begin{eqnarray}
\rho, ~\Omega=\left\{\begin{array}{lll} 
\rho_{1}\,,& \Omega_{1} & R\leq 1 \,, \\
\rho_{2}\,, & \Omega_{2} & R > 1 \,.
 \end{array}\right.
\end{eqnarray}
On either side of the discontinuity ($R=1$) the pressure distribution is determined by Eq.~(\ref{eq:balance}), up to an integration constant, which is chosen so that the pressure at the discontinuity is continuous, as required by its force balance. 
We fix this constant by setting the pressure at $R=0$.  We study both  Newtonian and  relativistic models, which are described in Table 1. For the relativistic runs, the speed does not exceed $0.9c$ and so the flows are only mildly relativistic. In all these models the instability criterion (\ref{eq:IC3}) is not satisfied on both sides of the discontinuity and hence the stability of the whole configuration is expected to be determine solely by the criterion (\ref{eq:IC2}) for the discontinuity.    
 \begin{table}
\centering
\begin{tabular}{rcclllrc} \hline\hline
Run &~ $\rho_{1}$ & ~$\rho_{2}$& ~ $\Omega_{1}$& ~$\Omega_{2}$&~$p(0)$ & ~$\Delta \hat{\Psi}$ &~ Stability \\ 
 \hline
 R1 & 1   & 1 &  0.9  &  0& 10 & $-2.00$& U\\
R2 & 1   & 1  & 0.9  & 0.45& 10 & $-1.76$& U \\
R3 & 1   & 2   & 0.45  & 0.45& 10& $0.51$& S\\
R4 & 2   & 1   & 0.45  &  0.45& 10& $-0.02$ & B\\
R5 & 10 & 1   & 0.30 & 0.45 & ~~0.1 & $-0.29$ & U \\
C1 & 1   & 1 &  2  &  1& 10 &$-1.19$ & U\\
C2 & 2   & 1   & 2  & 1& 10 &$-1. 55$& U\\
C3 & 1   & 2  & 2  & 1 &10& $-0.65$& U \\
C4 &1   &  2 & 1  &  1 & 10 &$0.68$ & S\\
C5 & 2   &  1 & 1  & 1  &10 &$ -0.65$&U\\
C6 & 2   &  1 & 1  & 2  & 10&$0.68$&S\\
C7 & 5   &  1 & 1  & 1.4  &10&$-0.87$&U\\
\hline  
\end{tabular}
\caption{Simulation models. 
The first column is the model name, C denotes the newtonian runs and R the relativistic ones. The second to fifth columns contain the density and angular momentum of the inner and outer fluid. The sixth is the pressure on the axis. The seventh is  $\Delta\hat{\Psi}=[\bar{\Psi}]/ \langle\Psi \rangle $, where $ \langle\Psi \rangle =\left(\Psi_{1}+\Psi_{2}\right)/2$. The last column shows if the initial configuration is found to be stable (S) or unstable (U). In the model R4 the growth rate of the instability is very slow; it is saturated at a rather small amplitude and eventually damped by numerical dissipation. For this reason it is marked as a borderline case (B).} 
\label{Table:1}
\end{table}

The simulations were carried out with the AMRVAC code. We used the HD module for the Newtonian case and the SRHD module for the relativistic one \citep{Keppens:2012,Porth:2014}. We  integrated the equations of ideal 
 fluids with  the adiabatic index $\gamma=5/3$ for the Newtonian models and $\gamma=4/3$ for the relativistic ones. The computational domain is $(R,z)\in(0,2)\times(0,2)$, with the periodic boundary conditions at $z=0$ and $z=2$. At $R=0$ we use the reflection boundary conditions and at $R=2$ we use symmetry conditions for $\rho$, $p$, $v_{\phi}$ and $v_z$ and the anti-symmetry condition for $v_R$. The initial equilibrium configuration is modified via a sinusoidal perturbation of the azimuthal velocity component in the vicinity of $R=1$, with wavenumber $k=5$ and  amplitude $10^{-3}$ of the local azimuthal velocity. 
 Most of the simulations were carried out on a uniform $400\times400$ grid. We have run some models with a higher resolution to check the numerical errors. In particular,  
the R4 and R5 models have been run with the double and quadruple resolutions -- the small $\Delta \hat{\Psi}$ of these models required to lower the numerical viscosity for the instabilities to develop. We also noticed the tendency for the higher resolution runs to produce finer features, suggesting a faster growth of modes with shorter wavelength. However, in this paper we focus on the instability criterion only and leave the growth rates issue to future studies.  

To track the fluids initially located at either side of the discontinuity, we used a passive tracer $\eta$ governed by the equation 
\be
\partial \left(\Gamma \rho \eta \right)/\partial t +\nabla \cdot \left(\Gamma \rho \eta \bm{v}\right)=0
\ee 
(in the non-relativistic case $\Gamma$ is set to unity). It is initialised so that $\eta=1$ for $R<1$ and $\eta =0$ for $R>1$.
\\
\begin{figure*}
\includegraphics[width=0.32\textwidth]{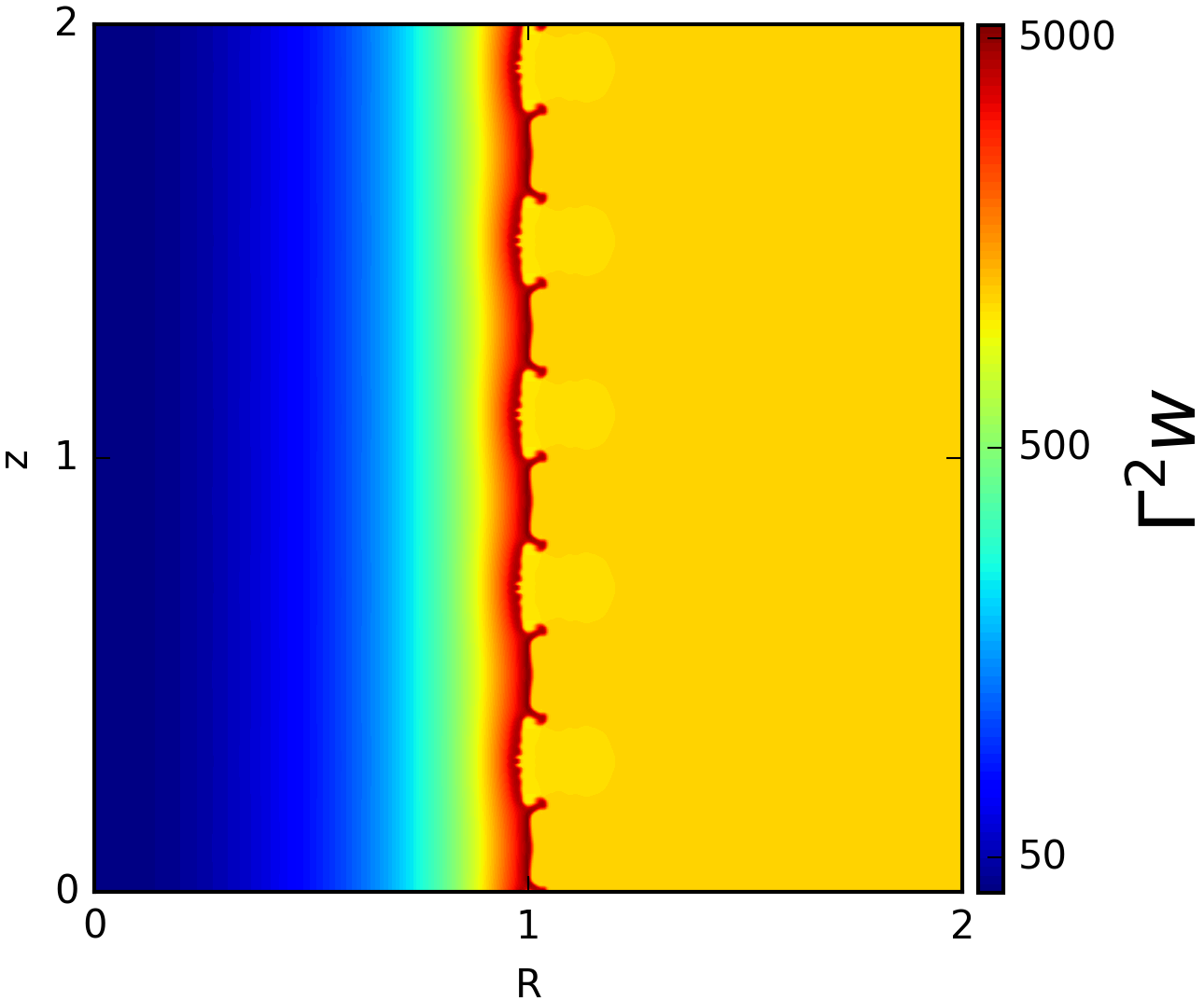}
\includegraphics[width=0.32\textwidth]{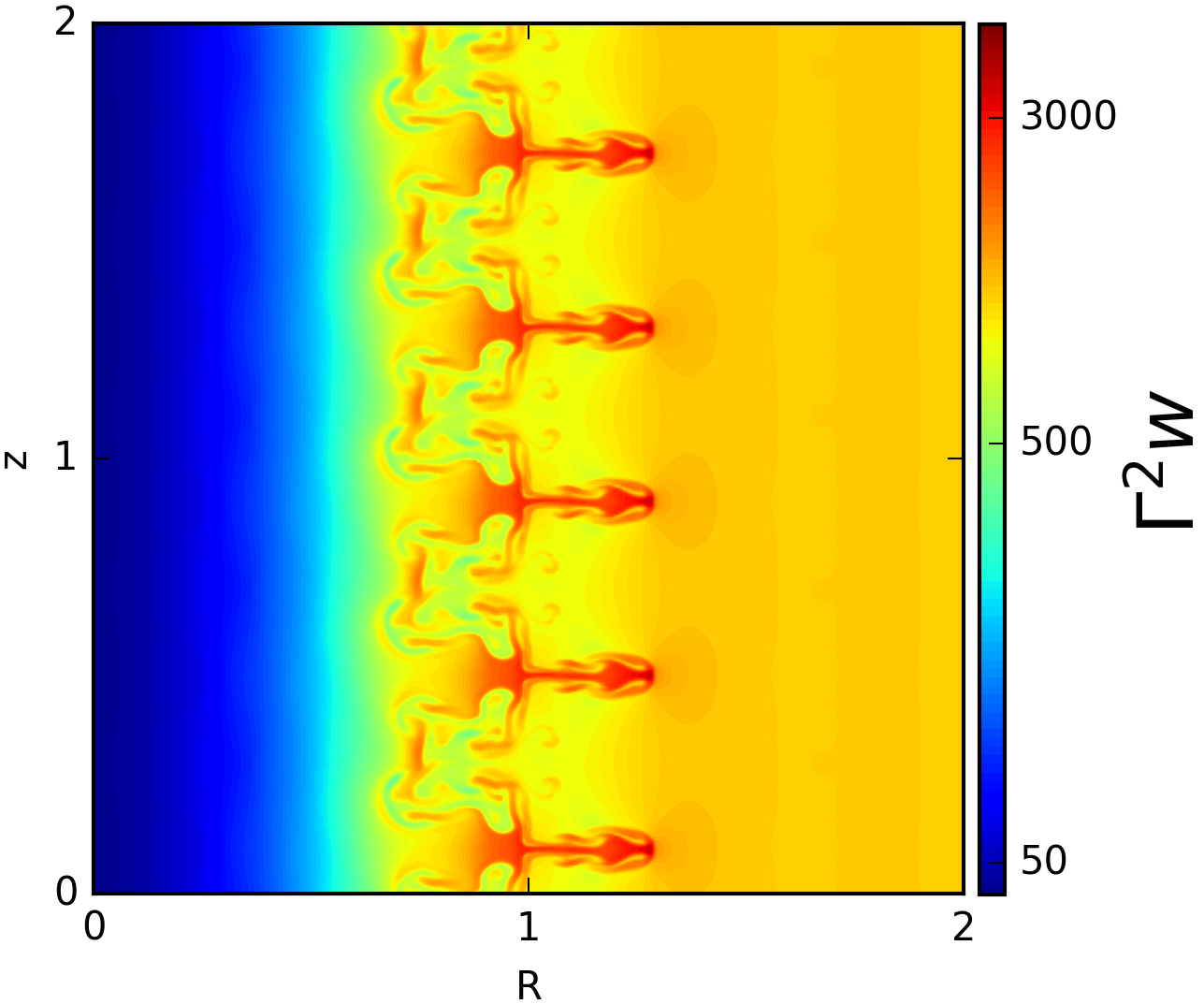}
\includegraphics[width=0.32\textwidth]{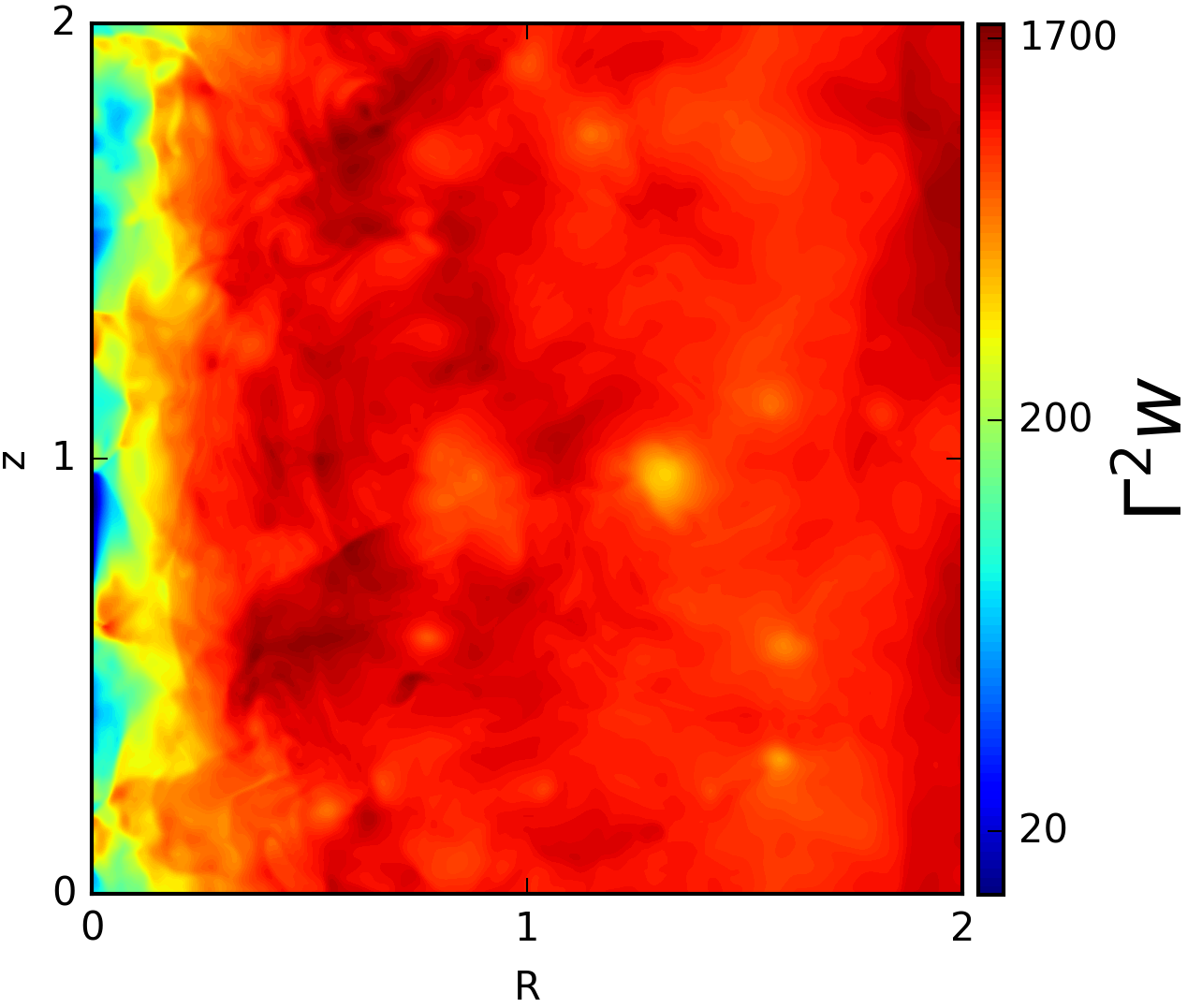}
\includegraphics[width=0.318\textwidth]{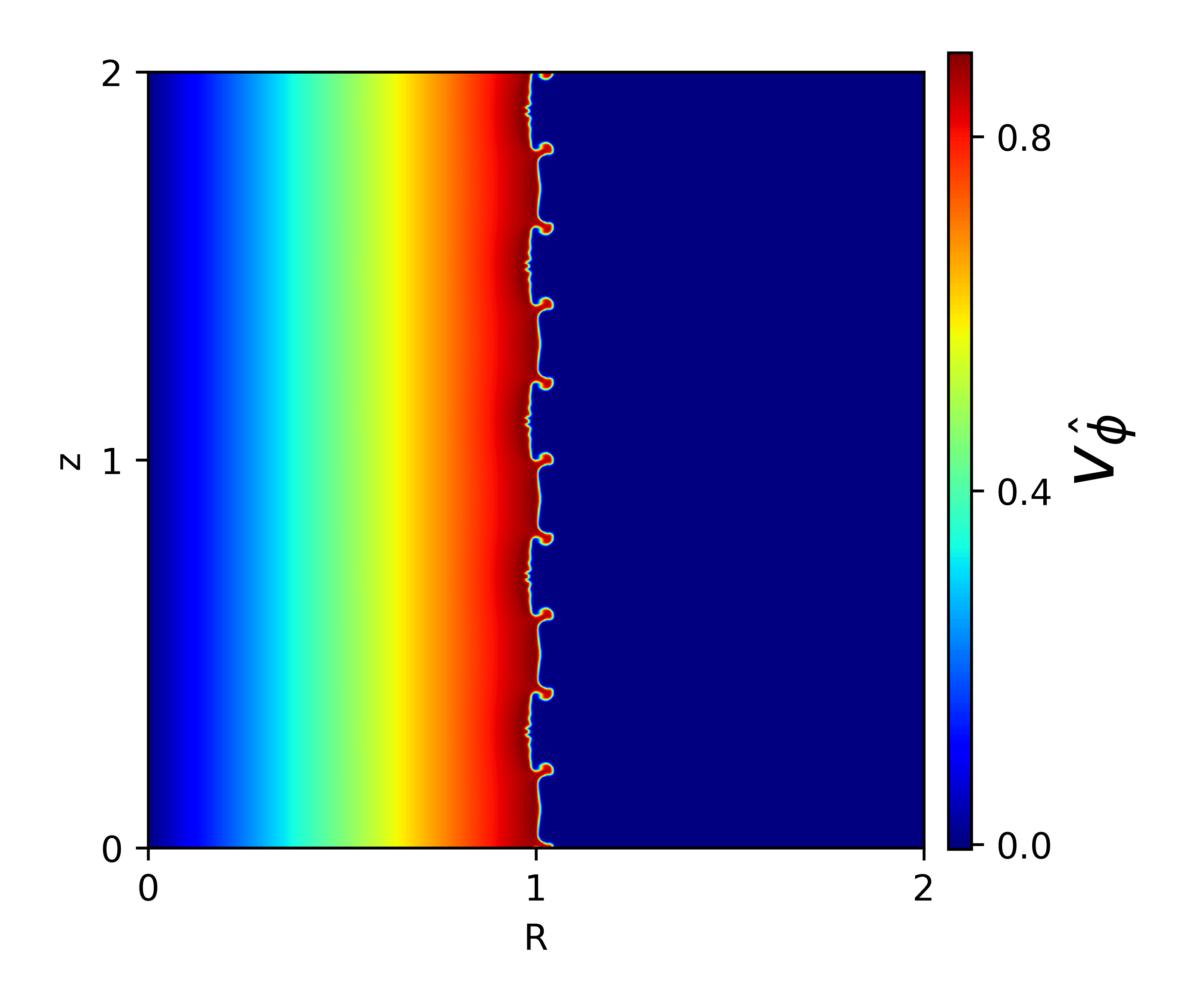}
\includegraphics[width=0.318\textwidth]{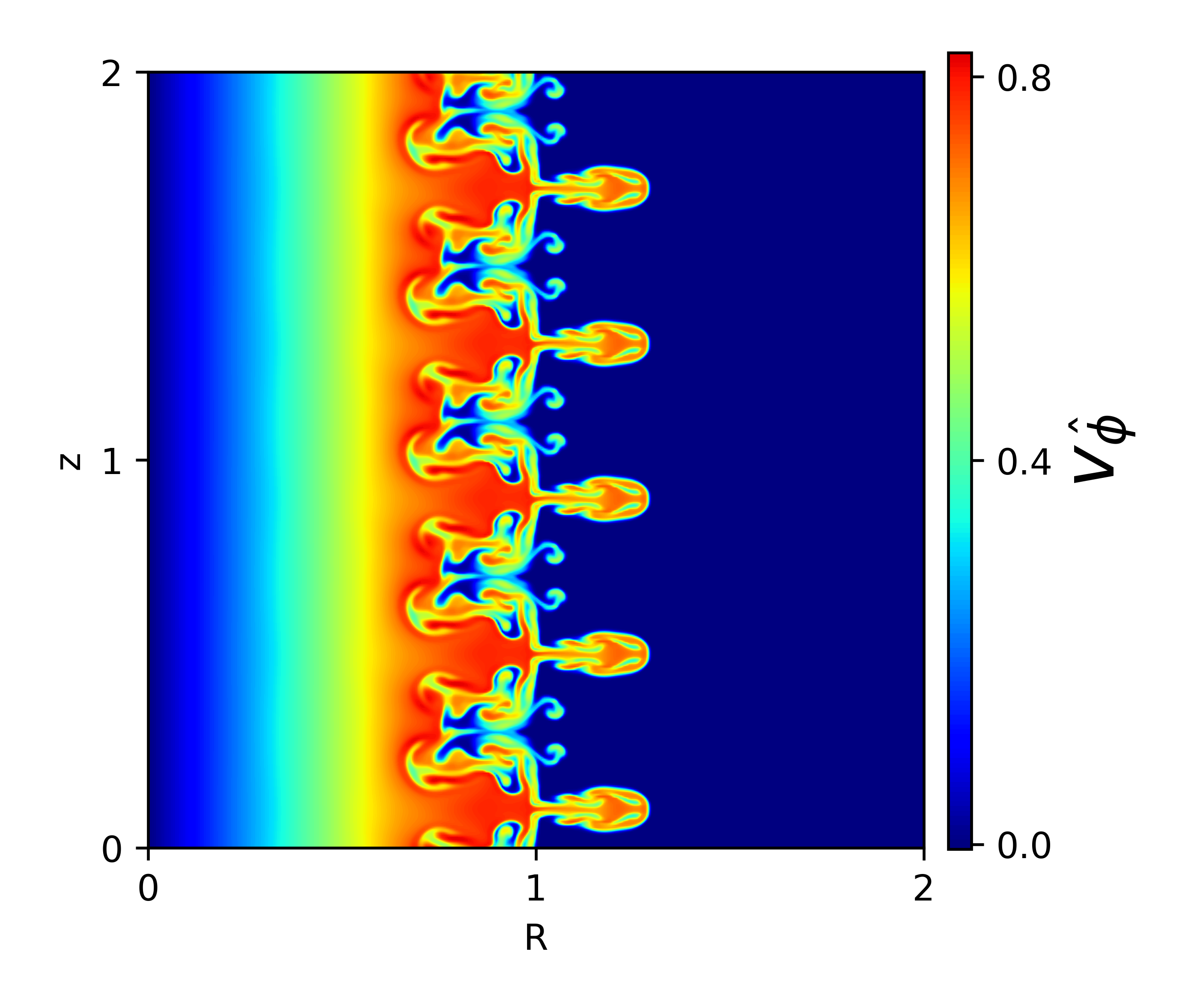}
\includegraphics[width=0.318\textwidth]{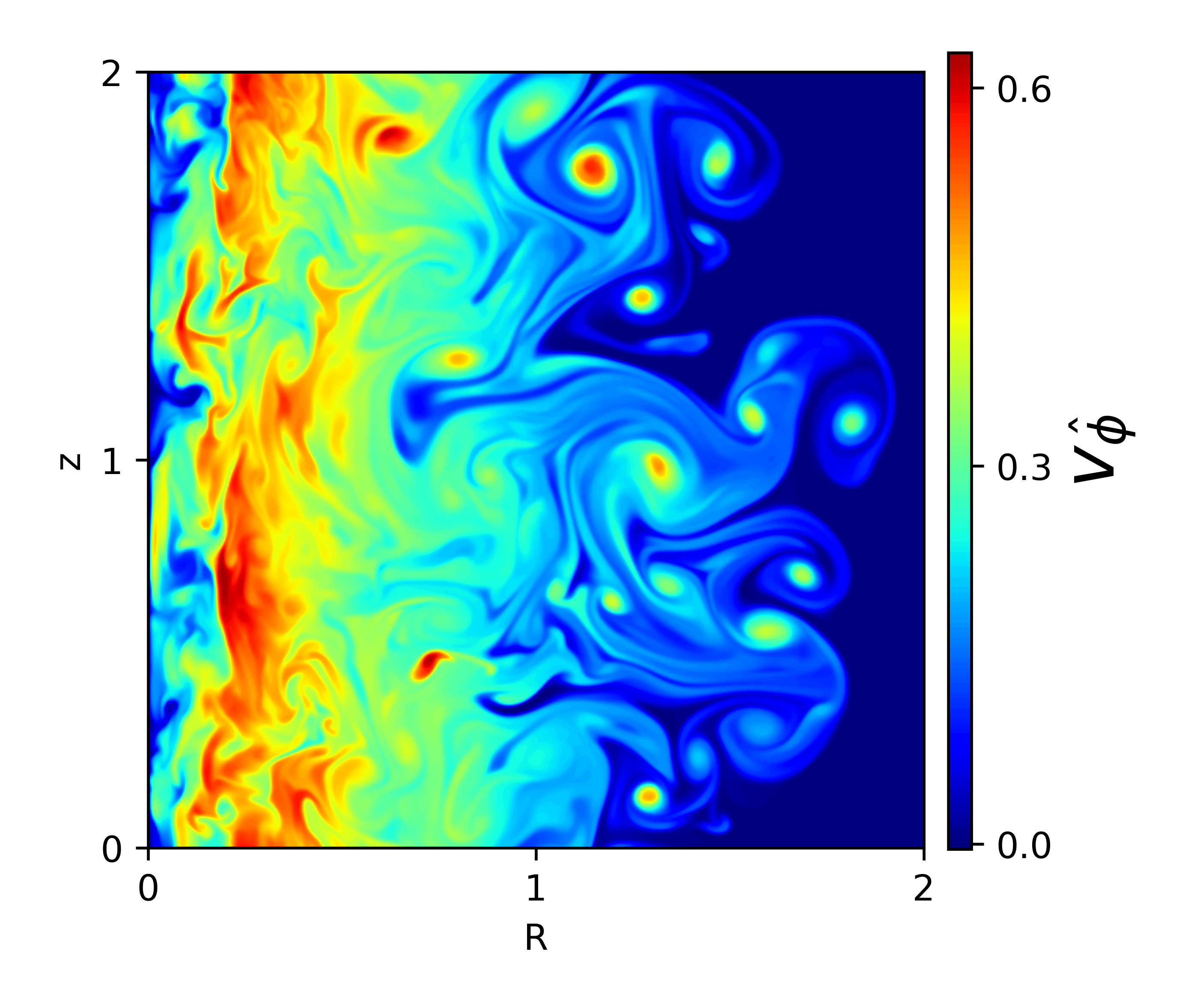}
\caption{The distributions of  inertial mass density $\rho_{in}=\Gamma^{2} w$ (top row) and  $v_{\hat{\phi}}=\Omega R$ (bottom row) for the model R1 at   $t=\pi/2\,, \pi\,, 4\pi$ (from left to right),  run at a resolution of $400^2$. }
\label{Fig:1}
\end{figure*}
\begin{figure*}
\includegraphics[width=0.318\textwidth]{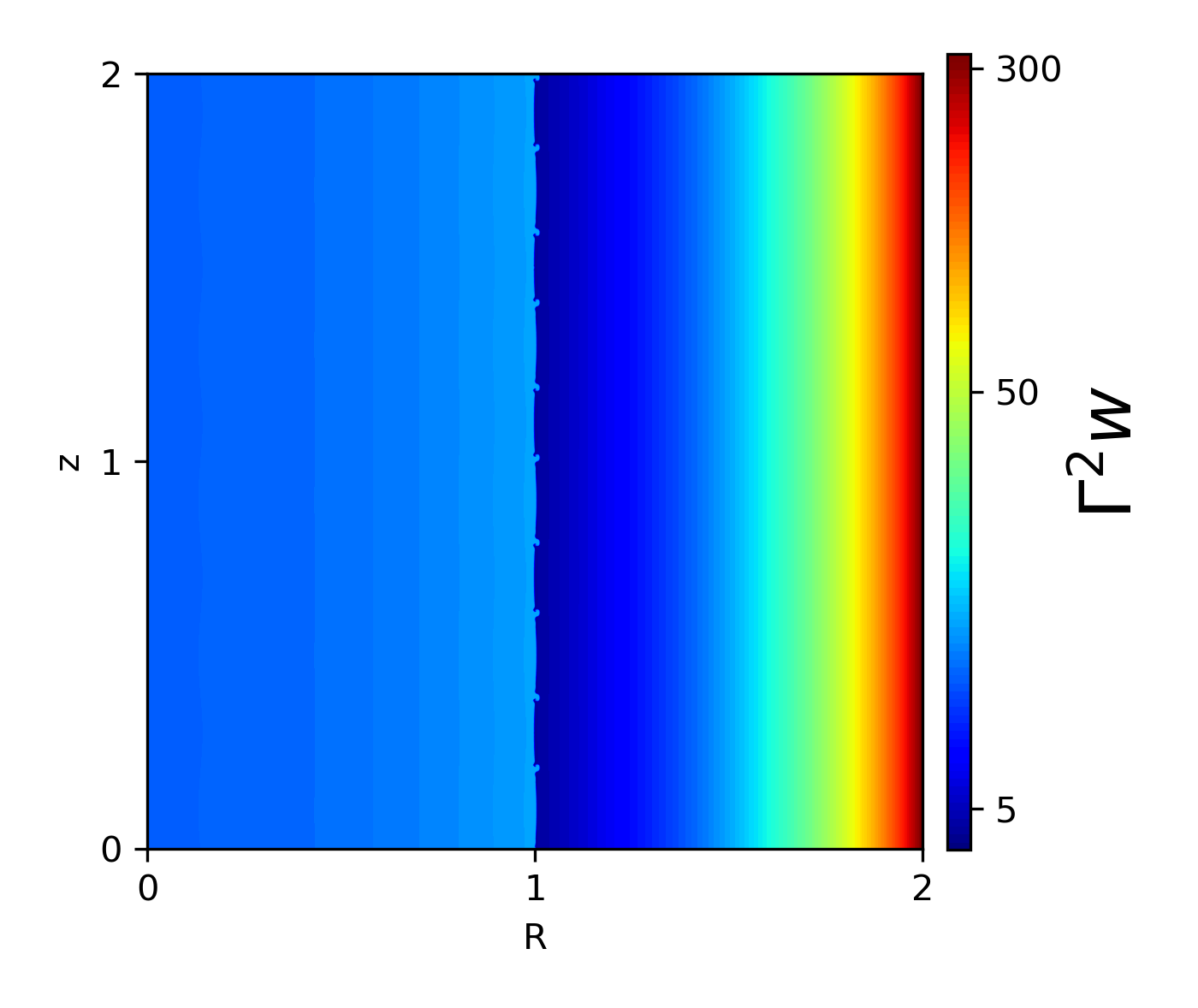}
\includegraphics[width=0.318\textwidth]{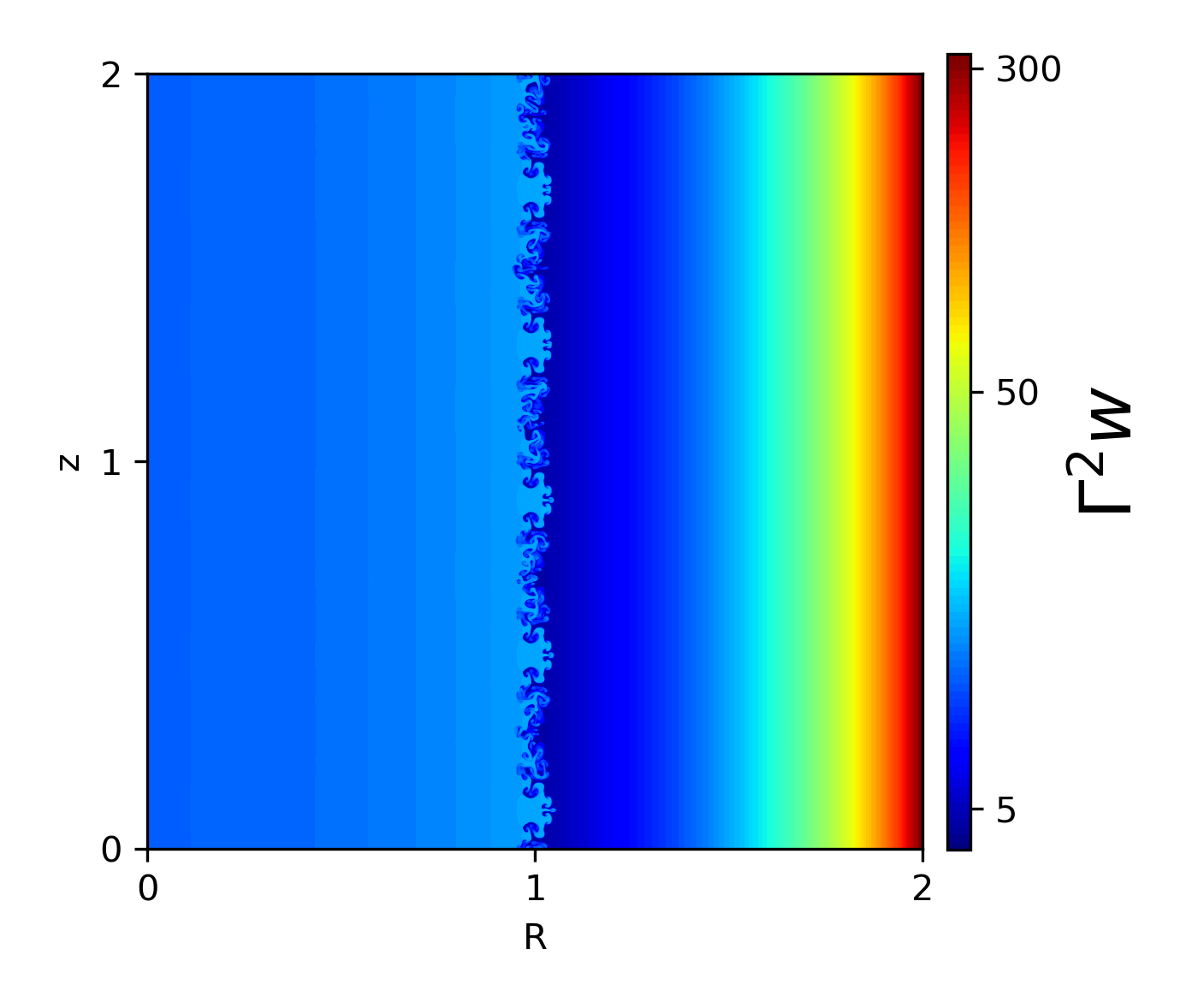}
\includegraphics[width=0.318\textwidth]{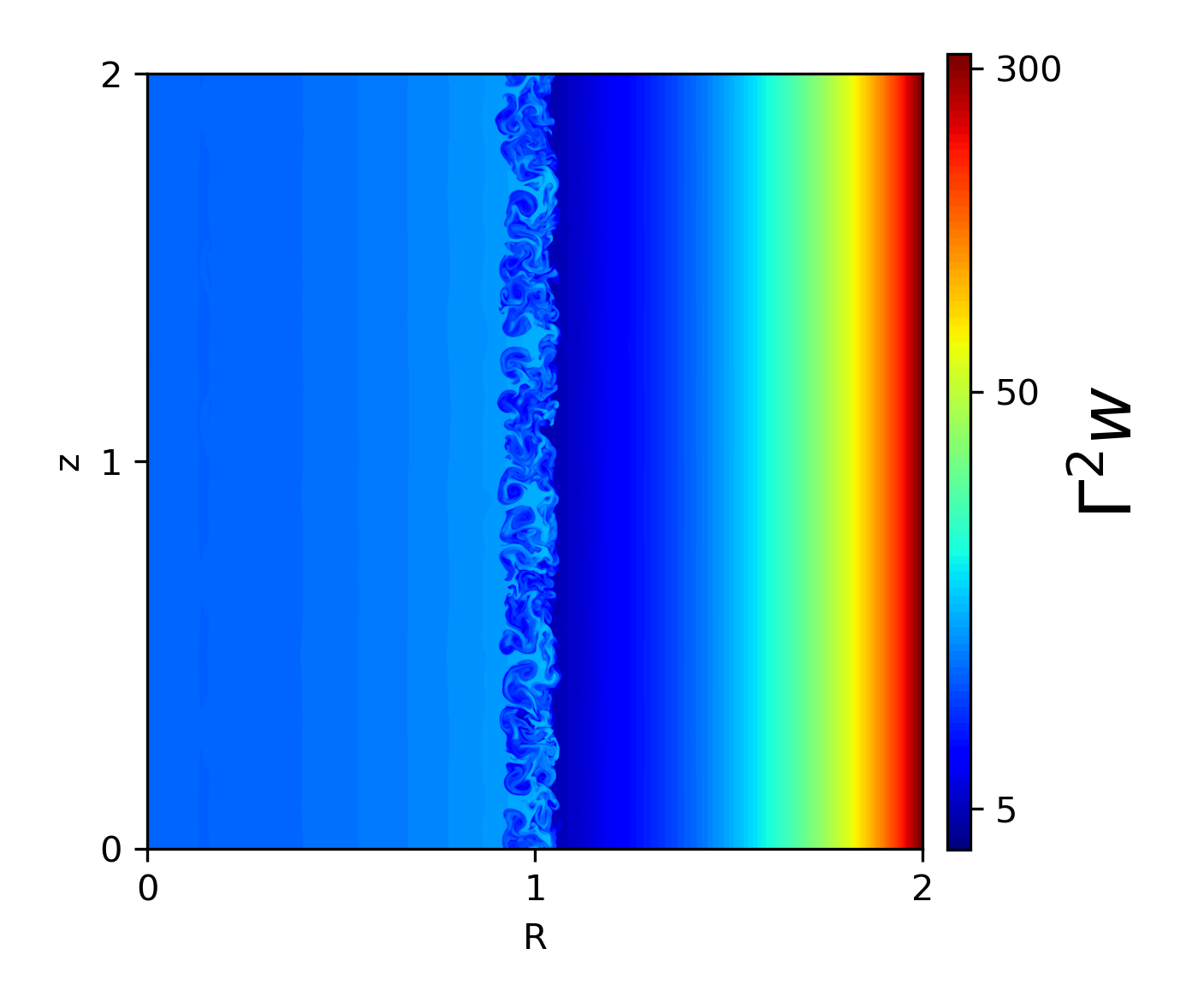}
\includegraphics[width=0.318\textwidth]{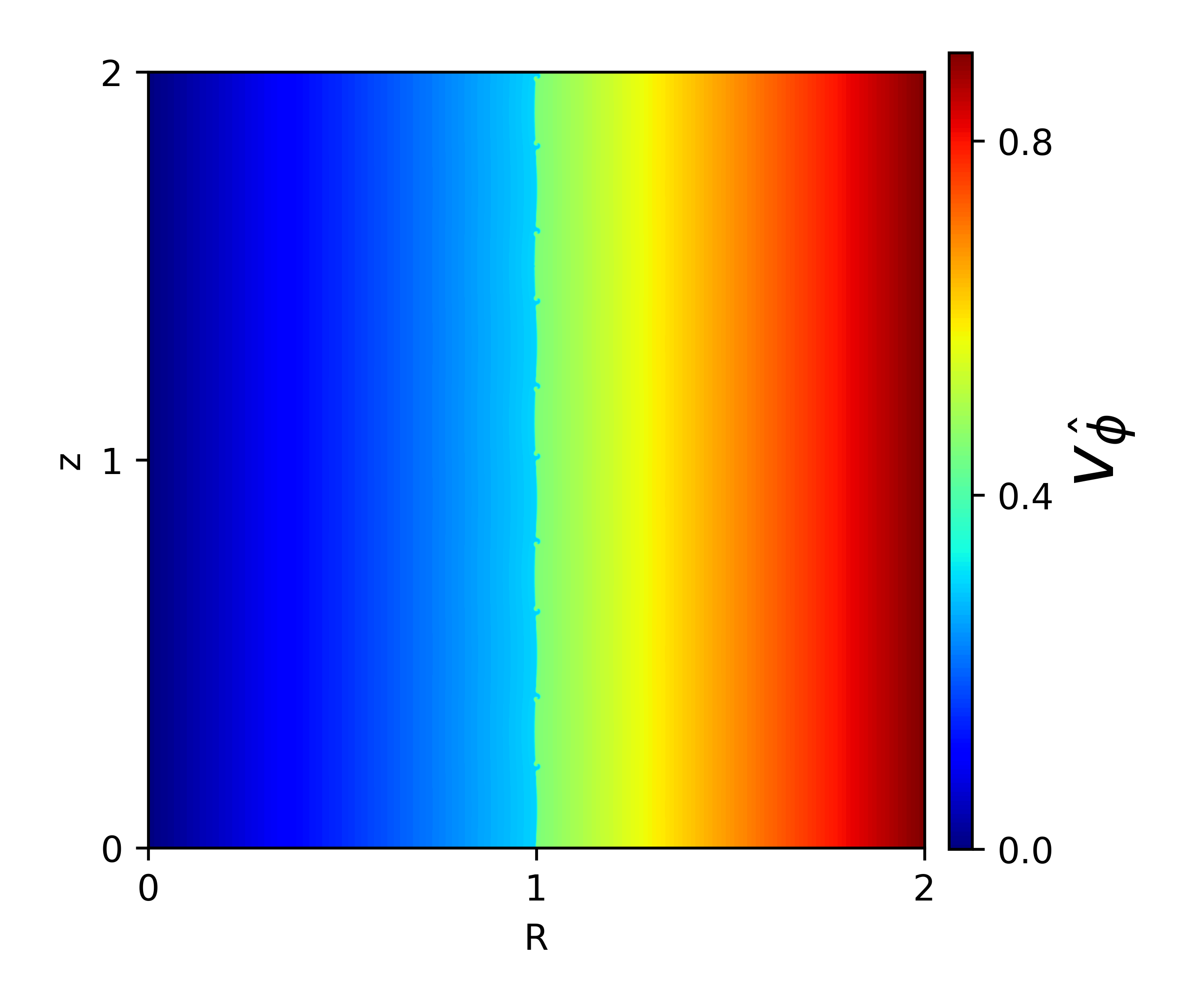}
\includegraphics[width=0.318\textwidth]{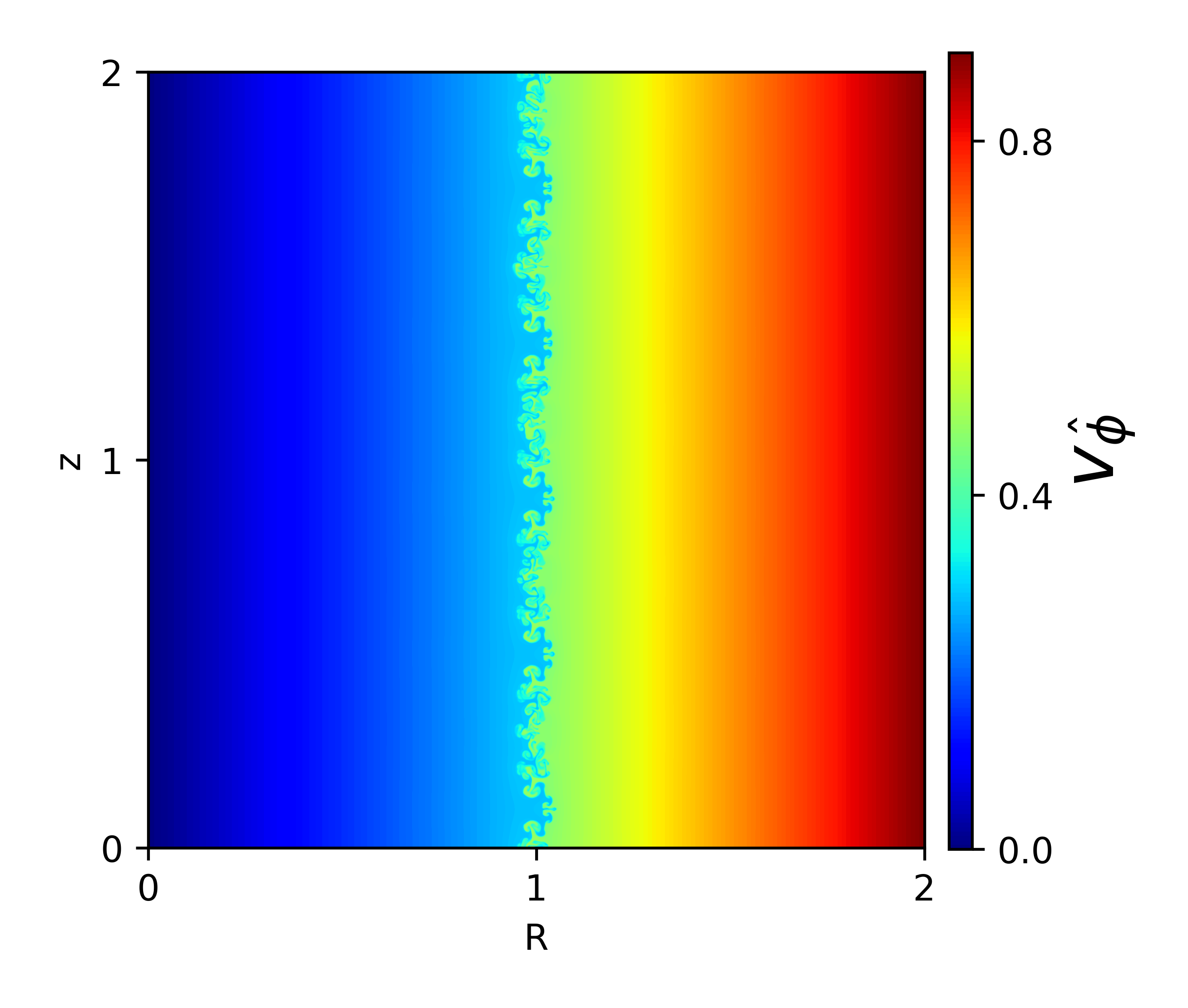}
\includegraphics[width=0.318\textwidth]{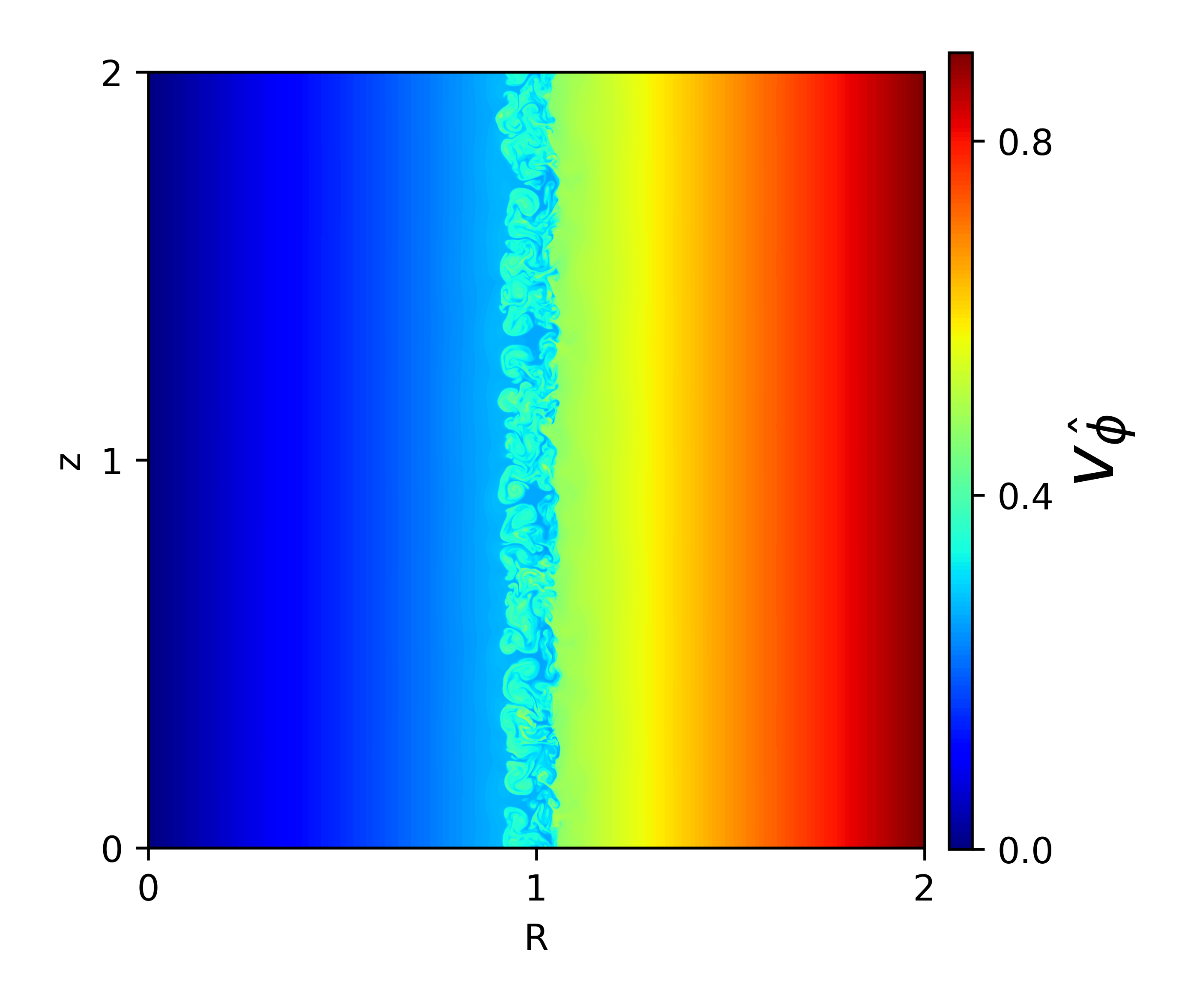}
\caption{The distributions of  inertial mass density $\rho_{in}=\Gamma^{2} w$ (top row) and $v_{\hat{\phi}}=\Omega R$  (bottom row) for the model R5 at   $t=2\pi \,, 4\pi\,, 8\pi$ (from left to right),  run at a resolution $1600^2$, allowing the development of finer structure features.}
\label{Fig:1a}
\end{figure*}
\begin{figure*}
\includegraphics[width=0.49\textwidth]{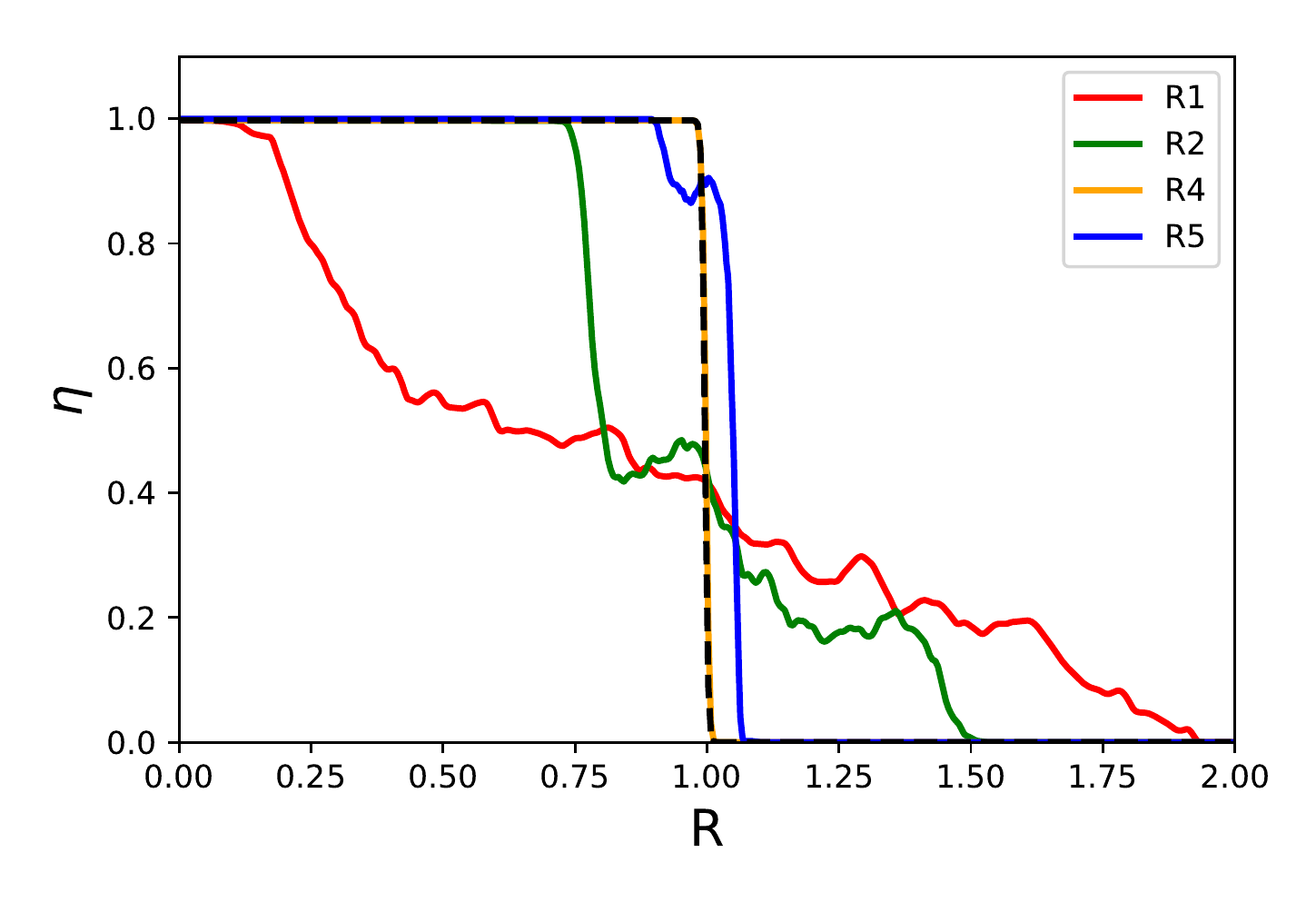}
\includegraphics[width=0.49\textwidth]{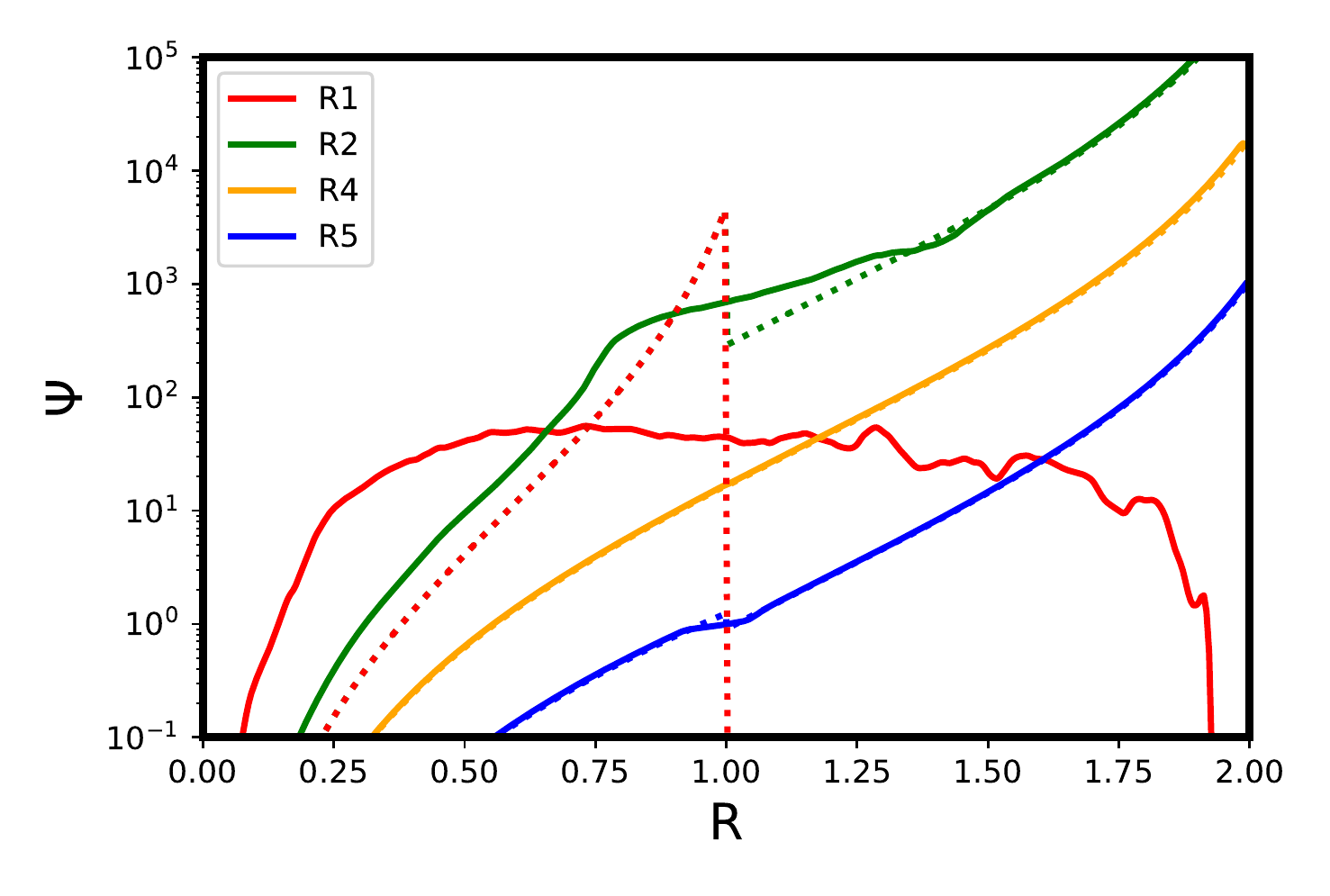}
\caption{Left panel: The $z$-averaged value of the passive tracer $\eta$ for unstable relativistic models (R1, R2, R4 and R5). The dashed line shows the initial distribution and the solid lines the final distributions (at $t=4\pi$, except for R5 where the final distribution is at $t=8\pi$).  The observed spread of the region with $0<\eta<1$ reflects the mixing of the two fluids.  Right panel: 
The $z$-averaged value of $\Psi=\Gamma^{2}w\Omega^{2}R^{4}$  for the same models. The dotted lines show the initial distributions (for R1 it drops to zero for $R> 1$) and the solid lines the final distributions.  The instability acts to remove the regions with the negative gradient of $\Psi$.  }
\label{Fig:2}
\end{figure*}

 As summarised in Table~\ref{Table:1}, the results of our simulations are in complete agreement with the generalised Rayleigh criterion (\ref{eq:IC2}), both in the Newtonian and relativistic regimes.   In particular, the relativistic models R1 and R2, where the initial density is uniform but the angular velocity on the outside of the discontinuity is smaller than on the inside and hence the specific angular momentum per unit mass  $l=\Omega R^2$ decreases with $R$, are unstable.  The R5 model, where the specific angular momentum increases with $R$, is also unstable, contrary to what one might have expected based on the original Rayleigh criterion but in agreement with the generalised one.  The relativistic models R3 and R4 have a uniform rotation and hence their initial configuration is analogous to the one of the Rayleigh-Taylor problem. The model R3 has a lighter fluid on the inside of the discontinuity and it is stable whereas the model R4 has a heavier fluid on the inside and it is unstable, as expected for RTI. 
 
 Among the unstable models,  R1 is totally disrupted by the end of the simulations (see Figure~\ref{Fig:1}), whereas R2 and R5 develop a turbulent layer around the discontinuity but remain mostly undisturbed elsewhere  (see Figure~\ref{Fig:1a}), and R4 shows very early saturation of the instability. This is illustrated in the left panel of figure~\ref{Fig:2} which shows the final distribution of the passive tracer.  These outcomes allow a simple interpretation: As the instability enters the nonlinear phase it begins to modify the spatial distribution of $\Psi$ by reducing the size of the region where the instability criterion is satisfied (see the right panel of Figure~\ref{Fig:2}). In the model R4, where the jump of $\Psi$ is very small, such a region is completely erased even before the instability reaches high amplitude.  In R2 and R5, where the jump is significantly larger, this occurs much later. Finally, in the model R1 there always exists a region where the instability criterion is satisfied because $\Psi=0$ in the undisturbed fluid outside of the layer.        

The same behaviour is observed for the Newtonian models (see C-models in Table~\ref{Table:1}). In particular, a higher jump of $\Psi$ leads to a more disturbed final solution. 
The initial configurations of C4 and C5 are analogous to that of the Rayleigh-Taylor problem and show that in this case the instability develops only when the inner fluid is heavier, as expected for RTI. In the model C3, the inner fluid is lighter but the instability still develops because the jump of the angular velocity ensures that the Rayleigh criterion for CFI is satisfied.  In the model C7, the instability develops even if the angular momentum per unit mass increases with $R$.  
    
The common feature of the nonlinear phase of CFI in all our unstable runs is the development by the inner fluid of elongated structures which penetrate  the outer fluid. These are reminiscent of the fingers associated with the normal Rayleigh-Taylor instability.   However, whereas RTI continues until the heavy and light fluids exchange their positions, which is accompanied by their mixing, CFI may terminate earlier, as soon as $\Psi(r)$ becomes a monotonically increasing function, and keep the most inner and outer sections of the initial configuration unaffected.

\section{Conclusions}

In this letter we have explored the CFI in axisymmetric rotating ideal relativistic and non-relativistic compressible flows.  We derived the generalised Rayleigh criterion for CFI for both the continuous and discontinuous flows and verified it via axisymmetric computer simulations for the discontinuous case. We consider this work as the first step in the studies of CFI in astrophysical jets. Even in the simplified case of rotating cylindrical flows, it remains to be seen if our generalised Rayleigh criterion holds for the continuous case.  Linear stability analysis is another important direction of study. 

The velocity shear in rotating flows may also be subject to the Kelvin-Helmholtz instability (KHI).  As a result of the axisymmetry, this instability is  suppressed in our simulations. The competition between the KHI and CFI is another important topic for future investigations. Finally, the astrophysical jets  possibly include strong magnetic field which may inhibit the growth of CFI and KHI modes and promote current-driven instabilities  \citep{Gourgouliatos:2012, Millas:2017}. Hence, the problem has to be expanded by including magnetic field. 

When it comes to astrophysical jets, it is important to go beyond the simple case of rotating cylindrical flow and explore the role of CFI in more realistic conditions. The first examples of such studies include the 3D simulations of rotating relativistic jets in flat spacetime \cite{Meliani:2007, Meliani:2009}\footnote{In these papers, the observed instability was interpreted as RTI.} and 3D simulations of relativistic jets undergoing reconfinement by external gas pressure (Gourgouliatos \& Komissarov, 2017).

\section*{Acknowledgements}
 We would like to thank Rainer Hollerbach for a very useful discussion of CFI in relation to AGN jets, as well as Zacharia Meliani, Christophe Sauty and Dimitris Millas for discussions on the distinction between CFI and RTI.  
This study was supported by STFC Grant No. ST/N000676/1. The numerical simulations were carried out on the STFC-funded DiRAC I UKMHD Science Consortia machine, hosted as part of and enabled through the ARC HPC resources and support team at the University of Leeds. 
\label{lastpage}

\bibliographystyle{mnras}
\bibliography{Bibtex}

\end{document}